\documentclass{IOS-Book-Article}     

\usepackage{mathptmx}
\usepackage{amsmath,amssymb}
\usepackage{cite}
\usepackage{color}
\usepackage{graphicx}
\usepackage{tikz}
\usepackage{url}
\usepackage{xspace}

\newcommand{\sR}{\mathbb{R}}
\newcommand{\dt}{\mathrm{d}t}
\newcommand{\clawpack}{{\sc Clawpack}\xspace}
\newcommand{\forestclaw}{ForestClaw\xspace}
\newcommand{\pforest}{\texttt{p4est}\xspace}
\newcommand{\manyclaw}{Manyclaw\xspace}

\newcommand{\ignore}[1]{}

\newcommand{\plotbox}[1]{#1}

\newcommand{\Fig}[1]{Figure~\ref{fig:#1}}
\newcommand{\Tab}[1]{Table~\ref{tab:#1}}

\begin{document}
\begin{frontmatter}          
%
\title{\forestclaw:
        Hybrid forest-of-octrees AMR for hyperbolic conservation laws}
\runningtitle{\forestclaw}

%
\author[A]{\fnms{Carsten} \snm{Burstedde}%
\thanks{Corresponding author.  E-mail: \texttt{burstedde@ins.uni-bonn.de}}},
\author[B]{\fnms{Donna} \snm{Calhoun}},
\author[C]{\fnms{Kyle} \snm{Mandli}} and
\author[C]{\fnms{Andy R.} \snm{Terrel}}
\runningauthor{C.\ Burstedde et al.}
\address[A]{Institut f\"ur Numerische Simulation, Universit\"at Bonn, Germany}
\address[B]{Boise State University, Idaho, USA}
\address[C]{Institute for Computational Engineering and Sciences,\\
The University of Texas at Austin, USA}

\begin{abstract}
We present a new hybrid paradigm for parallel adaptive mesh refinement (AMR)
that combines the scalability and lightweight architecture of tree-based AMR
with the computational efficiency of patch-based solvers for hyperbolic
conservation laws.  The key idea is to interpret each leaf of the AMR hierarchy
as one uniform compute patch in $\sR^d$ with $m^d$ degrees of freedom, where $m$ is
customarily between 8 and 32.  Thus, computation on each patch can be optimized
for speed, while we inherit the flexibility of adaptive meshes.  In our work we
choose to integrate with the \pforest AMR library since it allows us to compose
the mesh from multiple mapped octrees and enables the cubed sphere and other
nontrivial multiblock geometries.  We describe aspects of the parallel
implementation and close with scalings for both MPI-only and OpenMP/MPI hybrid
runs, where the largest MPI run executes on 16,384 CPU cores.
\end{abstract}


\end{frontmatter}


\section{Introduction}

With the advent of many-core chips such as GPUs and the MIC architecture
 comes the opportunity to sustain unprecedented rates of floating point
operations at comparably high integration density and low cost.  These
architectures, however, require careful structuring of the data layout and
memory access patterns to exhaust their multithreading and vectorization
capabilities.

Consequently, it is not clear a priori how to accelerate PDE solvers
that use adaptive mesh refinement.
Of course, it was realized early that it helps to aggregate degrees of
freedom (DOF) at the element level, as has been done with high-order
spectral element \cite{TufoFischer99}, low order continuous Galerkin methods that accumulate many elements simultaneously \cite{knepleyterrel:2013},
or discontinuous Galerkin \cite{HesthavenWarburton02}
methods.  GPU implementations of the latter have been proposed recently
\cite{KlocknerWarburtonBridgeEtAl09, BursteddeGhattasGurnisEtAl10}.
The finite volume method has typically been implemented using a single degree
of freedom per cell on
structured
\cite{ppm, clawpack}
or unstructured meshes
\cite{openfoam};
higher order methods have also been constructed by widening the stencil, for instance in WENO methods
\cite{Shu:2009bi}.


To facilitate hardware acceleration for parallel dynamic AMR, we
build upon the forest-of-octrees paradigm because of its low overhead
and proven scalability \cite{BursteddeWilcoxGhattas11}.  This approach
identifies each octree leaf with a mesh element.  The present work does not
construct a traditional high-order element but defines each element to be a
dense computational patch with $m^d$ DOFs.  In fact, this approach
resembles block-structured AMR \cite{be-ol:1984, be-co:1989,
ColellaGravesKeenEtAl07,
BerzinsLuitjensMengEtAl10} except that the
patches are not overlapping,
which enables us to capitalize on our
previous experience with scalable FE solvers for PDEs
\cite{BursteddeStadlerAlisicEtAl13}.  The \clawpack software
\cite{LeVeque97} provides a popular implementation of such a patch.
It has been designed to solve hyperbolic conservation laws
and successfully used in the context of block-structured AMR
\cite{be-le:1991, amrclaw, Berger:2011du}.

In this paper we describe our design for the coupling of forest-of-octrees AMR
with \clawpack at the leaf level.  We comment on challenges that arise in
enabling multiblock geometries and efficient parallelism and conclude with a range
of numerical examples that demonstrate the conceptual advantages.

\section{Design principles}

The starting point of our work is defined by the \pforest algorithms for
forest-of-octrees AMR on the one hand, and the \clawpack algorithms for the
numerical solution of hyperbolic conservation laws
on the other.  Both are specialized codes with the following characteristics:
\begin{center}
\begin{tabular}{l|l|l}
& \multicolumn{1}{c|}{\pforest} & \multicolumn{1}{c}{\clawpack} \\
\hline
subject & hexahedral nonconforming mesh &  hyperbolic PDE on $[0, 1]^d$ \\
toplevel unit & forest of octrees & patch of $m^d$ FV cells \\
atomic unit & octree leaf & one DOF in each cell \\
parallelization & MPI & threads (\manyclaw variant) \\
memory access & distributed & shared on each MPI rank \\
data type & integers & floating point values \\
language & C & Fortran 77 \\
\end{tabular}
\end{center}
Each leaf as the atomic unit of \pforest houses a toplevel unit of \clawpack.
The term cell is used to identify a single DOF within a \clawpack patch.  The
proposed 1:1 correspondence between a leaf and a patch thus combines two
previously disjoint models in a modular way:
\begin{enumerate}
\item We permit the reuse of existing, verified, and performant codes.
\item We preserve the separation between the mesh on one hand and the
discretization and solvers on the other.
\item The AMR metadata (\pforest:
under 1k bytes per octree, 8 bytes per MPI rank,
24 bytes per leaf. \forestclaw: $84 + 28d$ bytes per patch)
is insignificant compared to
the numerical data ($m^d$ floating point values per patch).

\item The resulting parallel programming model is a hybrid (often referred
to as MPI+X).  Only rank-local leaves/patches are stored and computed on.
\end{enumerate}

A particular feature of \forestclaw is that the generic handling of multiblock
geometries is inherited from \pforest, identifying each octree as a block.
Each block is understood as a reference unit cube with
its own geometric mapping.  The connectivity of the blocks can be created by
external hexahedral mesh generators, eliminating the need to encode it by hand.

A main challenge is presented by the fact that the patch neighborhood
is only known to \pforest.  This patch connectivity information needs
to be propagated to the numerical code in \forestclaw that implements
the interaction with neighbor patches via the use of a layer of (typically two)
ghost cells surrounding each patch.  To this end, we have designed an
interface that provides \ignore{read-only} access to the sequence of blocks, the list of
patches per-block, and a constant-time lookup of neighbor patches (and their
relative and, in general, nontrivial orientation between blocks).
Suitably informed by \pforest, \forestclaw stores only the patches local to
each MPI rank.

Mesh modification directives, such as adaptive refinement and coarsening, are
called from \forestclaw and executed as black-box operations inside \pforest.
The only information that flows into \pforest is a set of per-patch
refinement and coarsening flags, which are computed from the numerical state in
\forestclaw.  Our current algorithm requires neighbor patches to have a size
difference of at most a factor of two (the 2:1 balance condition).  This condition
can have nonlocal effects on the refinement pattern and is enforced by
special-purpose parallel algorithms in \pforest \cite{IsaacBursteddeGhattas12}.







\section{Patch-based numerics at the leaf level}


For hyperbolic problems, we integrate the solution on a single uniform patch,
containing $m^d$ cells, using the wave propagation algorithm described by R. J.
LeVeque \cite{LeVeque:1997eg} and implemented in \clawpack \cite{le:2002,
clawpack}.  We assume a single degree of freedom per cell and reconstruct a
piecewise constant solution to obtain left and right states at cell edges.  At
each edge, we solve Riemann problems to obtain left and right going waves
propagating at speeds determined from the solution to the Riemann problem.  For
scalar advection, the speed of each wave is simply the local advection speed at
the cell interface.  For non-linear problems and systems, an approximate
Riemann solver, such as a Roe solver \cite{roesolver}, is typically used.  Since
much of the physics of an application can be contained in the Riemann solver,
\forestclaw adopts \clawpack's interface to Riemann solvers effectively allowing
problems solved with \clawpack to be solvable in \forestclaw.  Wave limiters
are used to supress spurious oscillations resulting from truncation error in a typical
second order method.
\ignore{In order to
achieve second order accuracy wave limiters are used to suppress spurious
oscillations at sharp gradients in the flow.}

Data exchanges between neighboring patches are done via layers of
ghost cells extending the dimensions along the edges of each patch.
The interior edge values of a given patch overlap the ghost cell
region of a neighboring patch.  For the second order wave propagation
algorithm, two layers of ghost cells are sufficient.  This implies that
one layer of ghost patches is sufficient for $m \ge 4$.  Neighboring
patches at the same level of refinement simply copy their interior
edge values into a neighbors' ghost cells.  Neighboring fine grid
patches average their interior edge data to a coarser neighbor's ghost
cell values.  And neighboring coarse grid patches interpolate data
from their interior edge cells to their fine grid neighbor's ghost
cell values.  To avoid loss of conservation and the creation of
spurious extrema, we use a standard conservative, limited
interpolation scheme to interpolate values from the coarse grid to
fine grid coarse cells \cite{amrclaw, chombo}.  When sub-cycling, time accurate
data between coarse grids is used to fill in ghost cells for fine grids.
As we detail in the following section, this procedure can be extended
transparently to distributed parallelism by defining an abstract exchange
routine for ghost patch data.

Mesh refinement and coarsening requires interpolation from coarse grids
to newly created fine grids, and the averaging of fine grid data to a newly
created underlying coarse grid.  This operation is rank-local, analogous to the
general dynamic AMR procedures used in \pforest-based FE codes \cite[Fig.\
4]{BursteddeGhattasStadlerEtAl08}.

\section{Parallelization}

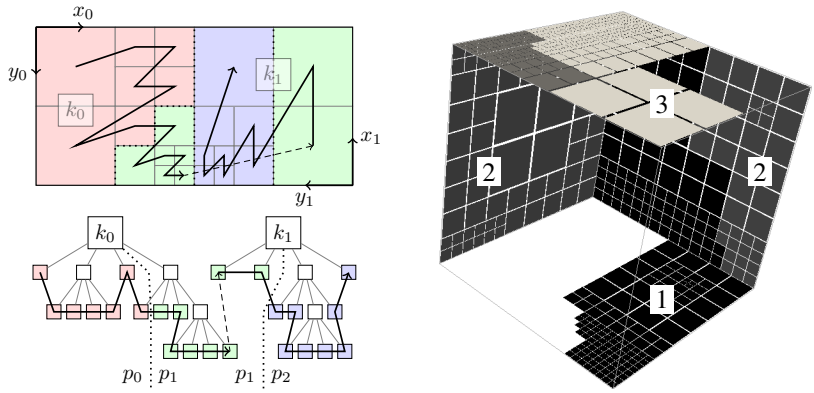
\begin{figure}
\begin{center}
\includegraphics[width=.4\columnwidth]{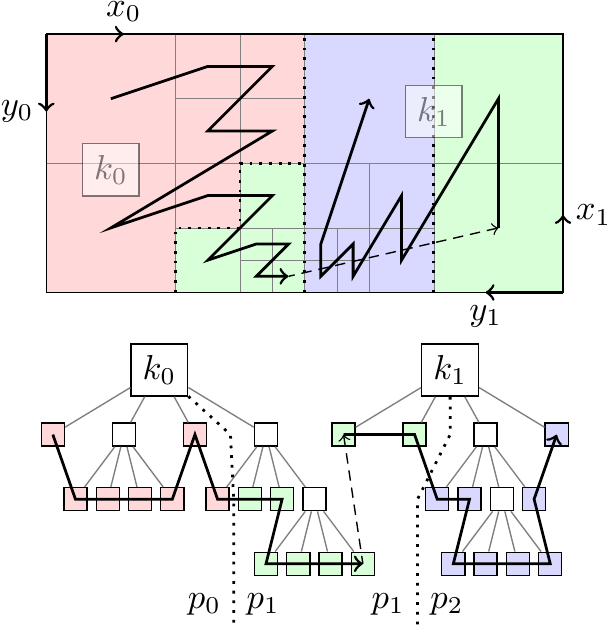}
\hspace{.05\columnwidth}
\begin{tikzpicture}[every node/.style={anchor=south west,inner sep=0pt},
                    x=1mm, y=1mm]
  \node at (0,0) {
    \includegraphics[width=.4\columnwidth]{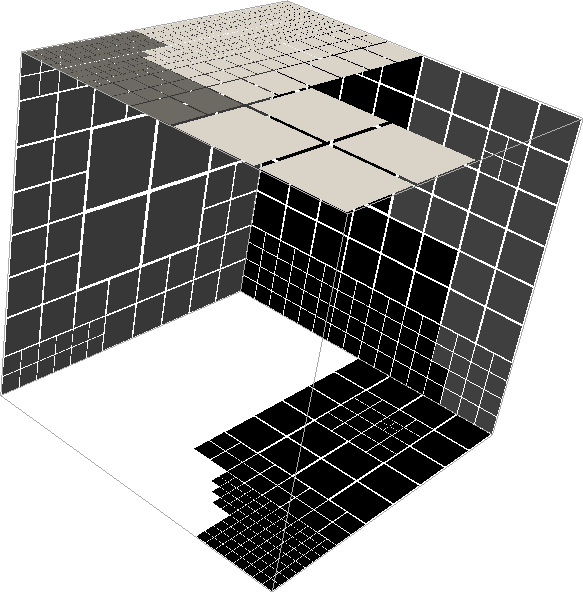}
  };
  \node [fill=white, inner sep=2pt] at (28, 10) {1};
  \node [fill=white, inner sep=2pt] at (5, 27) {2};
  \node [fill=white, inner sep=2pt] at (41, 27) {2};
  \node [fill=white, inner sep=2pt] at (28, 36) {3};
\end{tikzpicture}
\end{center}
\caption{Left: Forest of two quadtrees, partitioned among three MPI processes.
Each quadtree has its own coordinate orientation.  The situation in 3D
(octrees) is analogous.  Right: The leaves in a forest of six quadtrees that
serves as the computational domain for the cubed sphere.  An adhoc refinement
pattern has been 2:1 balanced in terms of neighbor sizes and partitioned
between five MPI processes (the three middle ones out of $0 \ldots 4$
are shown with a color scale from black to white).}
\label{fig:cubed3}
\end{figure}

The MPI layer is addressed from within \pforest and not exposed to the
\forestclaw code.  The \ignore{order of leaves} leaf ordering
is maintained in \pforest according to a
space filling curve.  Each MPI rank has a local view on its own partition,
augmented where necessary with information about one layer of ghost leaves
(see \Fig{cubed3}).

\forestclaw uses iterators over all rank-local leaves to execute numerical
tasks one patch at a time, optionally restricted to a given refinement level.
Random access to patches is possible and used when dereferencing the results
from
neighbor lookups.  Looping over the patches in the order
prescribed by the forest
and
accessing neighbors only relative to the current patch leads to a high
percentage of cache reuse
\cite{BursteddeBurtscherGhattasEtAl09}.

When \forestclaw accesses neighbor patches, they can be on the same or a
different block.  In the latter case, coordinate transformations are carried
out.  The structure of \forestclaw is oblivious to the fact that it only has a
local view of the distributed mesh and data which relieves the developer from
programming to the MPI interface.

With MPI parallelism, neighbor patches can be either local or assigned to a
different process (ghost patches).  Since this must make no difference
numerically, we need to ensure that the values of parallel neighbor patches are current
whenever they are accessed.  To this end, we allocate storage for a layer of
ghost patches in \forestclaw, which we can pass to a general-purpose \pforest
routine that communicates local leaf data to all processes that view a
particular leaf as a ghost.  If we call this routine before we go through the
local neighbor interactions, we can handle ghost values implicitly by neighbor
lookups without querying if they are local or remote.


In the context of finite-element or finite-difference methods, there should be
one such parallel data exchange per time step for a global value of the time
step length $\dt$, or one exchange per discretization level per time step if
$\dt$ is chosen per-patch depending on its size (this is sometimes called
sub-cycling).  In the tradition of block-structured AMR codes, interaction
between neighbor patches is done in a hierarchy from coarse to fine levels, and
then correction factors are propagated back from fine to coarse levels, requiring
an exchange at each level.  With sub-cycling, this entails a recursion with an
operation count that is exponential in the difference between the largest and
the smallest refinement level, with the benefit that $\dt$ matches the CFL
condition at each level.  While a large number of exchanges per time step can
present a scaling bottleneck due to the inherent synchronization and latency
losses, the expectation is still that the time to solution improves when
switching from uniform to adaptive meshes, since we are computing with fewer
patches, and improves again when enabling sub-cycling, since we take larger
time steps on the coarser levels.

The space-filling curve paradigm allows for lightweight repartitioning
algorithms.  Even uniform meshes benefit from this approach since the number of
processors does not need to be commensurable with the number of patches in each
space dimension.  For adaptive meshes, we repartiton after every refinement
operation and transfer the numerical data accordingly; see \cite[Fig.\
4]{BursteddeGhattasStadlerEtAl08} for the overall procedure.  When using
sub-cycling, we have the option to assign a weight to each patch based on its
level, determined by the expected number of sub-cycles per coarse-level time
step, and partition to equidistribute the cumulative weight between the
processors.  Currently, we are not enforcing a load balance that is attained
separately for each level, as it is done for example in Chombo
\cite{ColellaGravesKeenEtAl07} or recent geometric adaptive multigrid schemes
\cite{SundarBirosBursteddeEtAl12}.

The threaded parallelism over the degrees of freedom of a patch can be handled
by \forestclaw alone without the need to involve \pforest.  For instance, a
many-core implementation, such as \manyclaw \cite{manyclaw}, can be used for
the integration of the hyperbolic system on a leaf-patch thereby allowing for
hybrid parallelism.  The design of leaf-patches then can enable efficient
management of data local to many-core architectures (for example the Xeon Phi
coprocessor) and the host.






\section{Numerical results}

\begin{figure}
\begin{center}
\plotbox{\includegraphics[width=.35\columnwidth]{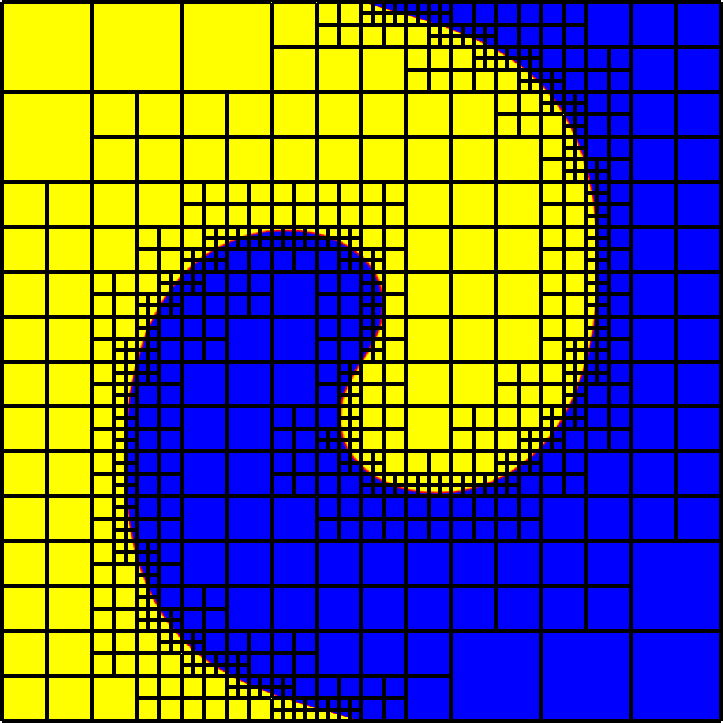}}
\hspace{.05\columnwidth}
\plotbox{\includegraphics[width=.35\columnwidth]{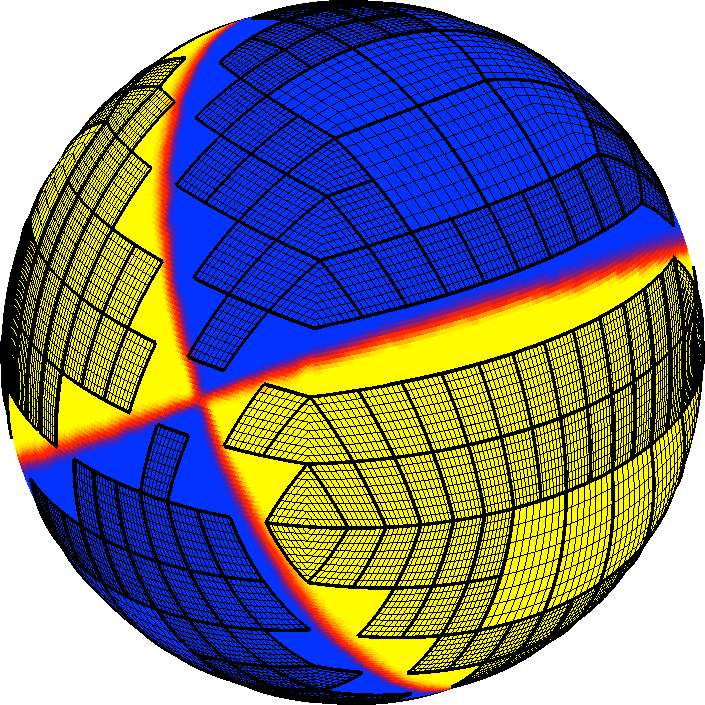}}
\end{center}
\caption{Numerical results for solving the advection equation.  Left:  Unit
square (single quadtree) with a twisting velocity field.  Right: Spherical ball
with a rotational velocity field constructed from two mapped quadtrees.  In
both cases the concentration is color-coded with a sharp gradient shown in red.
The adaptive mesh refinement follows the location of the gradient (the patches
are not shown where they become too fine for display).  Here we use $m = 8$
cells per \clawpack patch.}
\label{fig:results2d}
\end{figure}

We provide two examples that demonstrate the \forestclaw code.
The numerical results to date have been designed to verify that
interface layer between \pforest and \forestclaw is sufficiently
flexible and robust.  Of particular importance was ensuring that all
ghost cell transfers (including averaging and
interpolation) are implemented correctly.
The basic \clawpack algorithm and corresponding code are
thoroughly tested and need no further verification in our context.

In both sets of numerical results, we solve a scalar advection
equation,
\begin{equation}
q_t + ({\bf u}\cdot \nabla) q = 0 ,
\end{equation}
where the velocity field ${\bf u} = (u(\xi,\eta,t), v(\xi,\eta,t)$ is
a prescribed function of computational coordinates.  The relevant
numerical parameters that are set in each case include the size of the
patch on each leaf ($m=8$, $16$, or $32$), and the minimum and maximum
refinement levels, which in turn fix minimum and maximum effective mesh
resolutions.

In both examples, a patch is tagged for refinement if the
difference between its maximum and minimum values exceeds a prescribed
threshold.  A family of $2^d$ patches is coarsened to a parent patch
if this patch would not meet the criteria for refinement.

\begin{description}
\item{\bf Example 1:} An initial concentration field $q$
is set to 0 in the left half of a computational square and to 1 in the
right half.  A time dependent flow field is prescribed that distorts
the interface between the 0 and 1 concentration values.  
\Fig{results2d} shows the results at an intermediate time step, where
the minimum and maximum levels of refinement are set to 3 and 6, respectively.
This results in a minimum resolution of $64 \times 64$ and a maximum
resolution of $512 \times 512$.


\item{\bf Example 2:} We demonstrate the
multiblock functionality of \forestclaw by considering flow on a
sphere.  The sphere mapping we use consists of two quadtrees, each
defined in the computational space $[-1,1] \times [-1,1]$
\cite{ca-he-le:2008, be-ca-he-le:2009}.  Each quadtree is mapped to cover one
hemisphere.  The initial condition is $q = 0$ in one half of
the sphere, and $q = 1$ in the other half, where the halves are not
necessarily aligned with the equator of the mapping.  The flow field
${u}$ simulates rigid body rotation.  We show the results at an intermediate
time in \Fig{results2d}.
\end{description}

\begin{figure}
\begin{center}
  \includegraphics[angle=-90,width=.75\columnwidth]{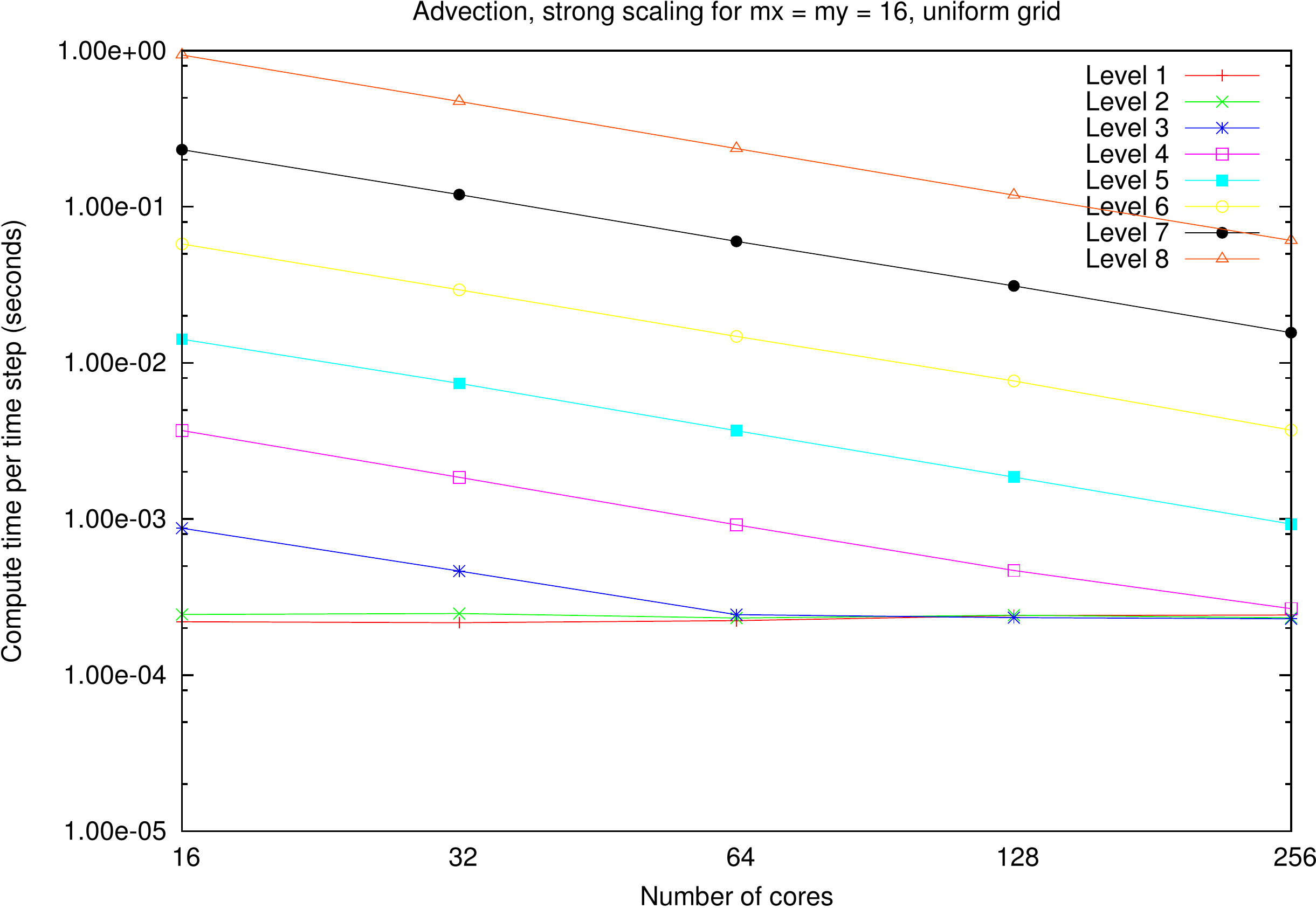}
  \\[1ex]
  \includegraphics[angle=-90,width=.75\columnwidth]{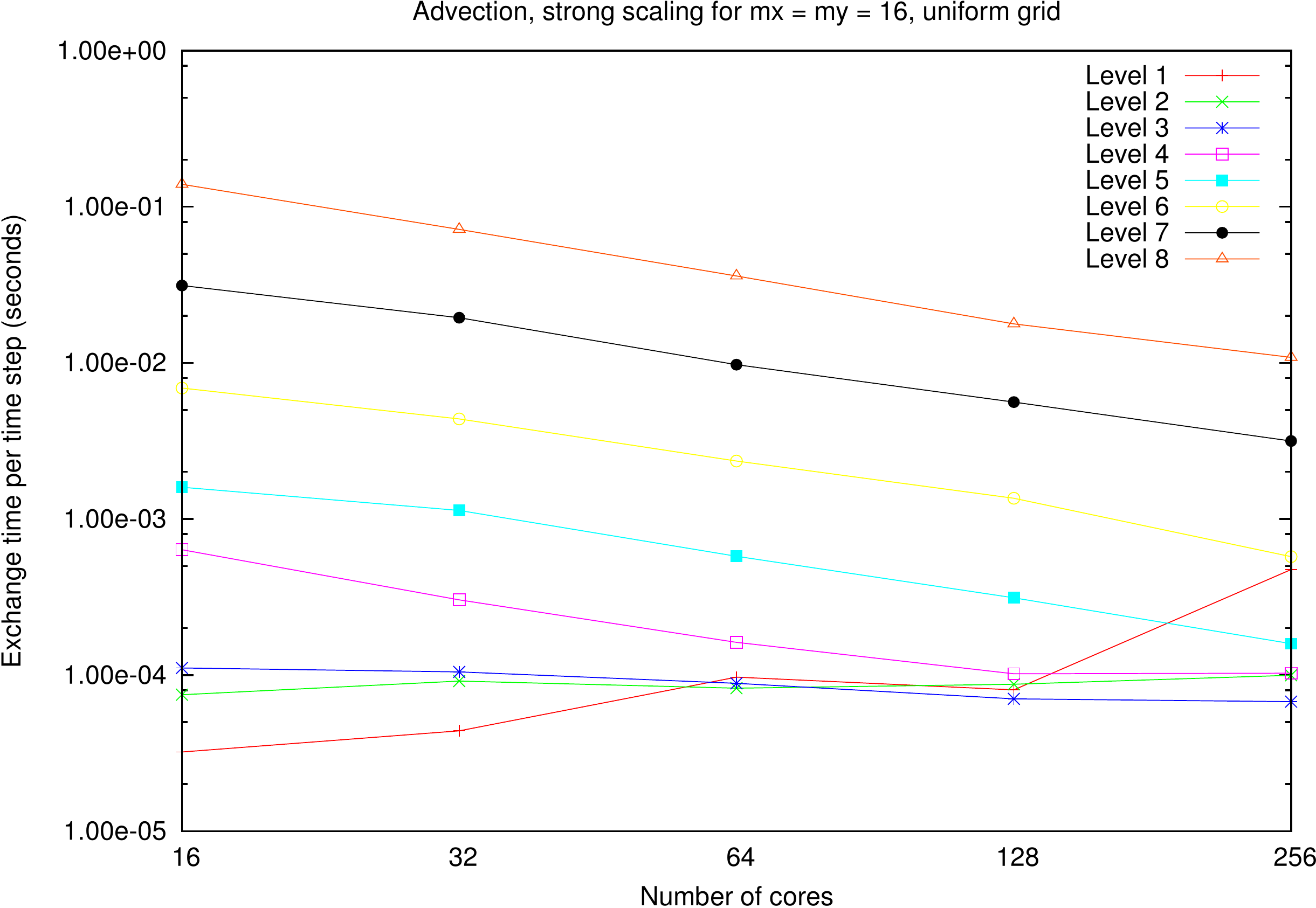}
\end{center}
\caption{Strong scaling of MPI parallelism for advection on a
  uniformly refined mesh.  Top: run time of time integration.  Bottom:
  time spent in local and parallel neighbor exchanges.  The number of
  patches is $2^{2 \ell}$ at level $\ell$.  The flat lines for the
  smaller runs are caused by having fewer patches than MPI processes,
  leaving some of the processors idle.  For each data point we use the
  time required by the slowest processor to verify that the load is
  balanced equally.
}
\label{fig:uniscale}
\end{figure}%
We begin our analysis of Example 1 with uniformly-refined mesh experiments run on TACC's
Stampede supercomputer at different levels of resolution.
We use 1 to 16 MPI processes on one 16-core compute node, obtaining a strong
scaling efficiency for the time integration between 97\%--103\% depending on
the resolution (from levels 1 and 8).  Then, we examine processor counts from
16 to 256 using 16 cores per node.  The scaling behavior remains linear for
the time integration itself, and mostly linear for the local and parallel
neighbor exchanges which require about 10\% of the time integration run time;
see \Fig{uniscale}.

\begin{figure}
\begin{center}
  \includegraphics[angle=-90,width=.75\columnwidth]{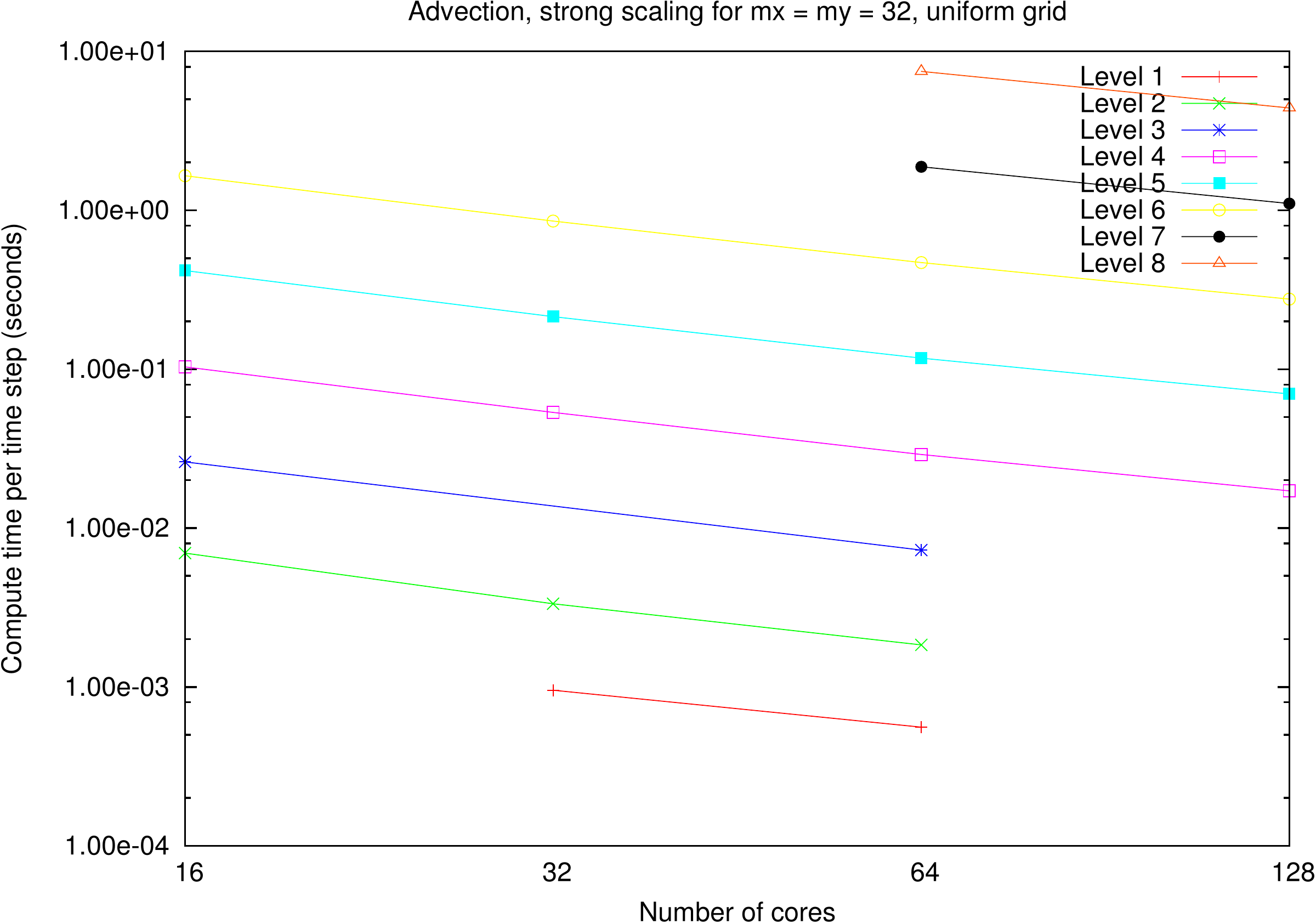}
\end{center}
\caption{Strong scaling of MPI/OpenMP parallelism for advection on a
  uniformly refined mesh with 16 threads per node.
}
\label{fig:ompscale}
\end{figure}%

In \Fig{ompscale}, we show multi-node calculations using 16 OpenMP
threads per node executed by the
\manyclaw set of patch-based PDE solvers.


We have examined weak scalability as well by using four times as many MPI
processes
for each level increase, from level 6 at 64 cores to level 11 at 16,384.  Up to
4,096 cores, the exchange times stay consistently below 16\% of the time
integration and the time in seconds per time integration step varies between
.0578 and .0602, which yields a parallel efficiency of 96\%.  For the
data point at 16,384 cores, the efficiency is still 82\%.

\begin{table}
  \begin{center}
    \begin{tabular}{cccc|rr}
      \multicolumn{4}{c|}{strategy} & \multicolumn{2}{c}{wall clock time} \\
      mesh & remesh & partition & time step & $P = 16$ & $P = 256$ \\
      \hline
      uniform & none       & by count  & global   & 3961.  & 256.  \\
      \hline
          AMR & every step & by count  & global   &  252.  &  54.6 \\
          AMR & every 4    & by count  & global   &  178.  &  39.7 \\
      \hline
          AMR & every step & by count  & subcycle &   99.9 &  17.3 \\
          AMR & every 4    & by count  & subcycle &   87.2 &  14.0 \\
      \hline
          AMR & every step & by weight & subcycle &   95.7 &  18.2 \\
          AMR & every 4    & by weight & subcycle &   84.4 & 14.2 \\
      \hline
    \end{tabular}
  \end{center}
\caption{Comparison of different meshing strategies for the advection example
  with a fixed maximum level 8, run on 16 and 256 cores, respectively.  The
  wall clock time includes the whole run of the program beginning to end (file
  I/O disabled).  The adaptive runs converge to a minimum level of 3.}
\label{tab:swirlwall}
\end{table}
Next, we examine how we can reduce the overall wall clock time of a simulation
by switching from uniformly refined meshes to adaptive refinement with the same maximum level,
effectively coarsening where high resolution is not needed.  For an example
with a sharp front as depicted in \Fig{results2d} we expect considerable
savings by AMR, which we confirm in \Tab{swirlwall}.  The wall clock times can
be reduced by up to a factor of 50 by combining AMR with subcycling.  The
adaptive runs are still  load balanced overall, which we infer from
the linear scaling of the time integration alone (not shown) and from the fact
that the weighted partition does not influence the run times.  There is the
tradeoff however, at least in the current implementation, that the adaptive
runs exhibit reduced scalability associated with the ghost patch exchanges
described earlier.  Presently, ghost patch exchanges are made across all levels, even
if only two levels are involved in the exchange.  One enhancement of the code that is
currently being developed is to rewrite the time stepping loop with the
objective of minimizing the parallel neighbor exchanges.

\begin{figure}
  \begin{center}
    \hspace{.1\columnwidth}
    \begin{minipage}[c]{.30\columnwidth}
      \includegraphics[width=\columnwidth]{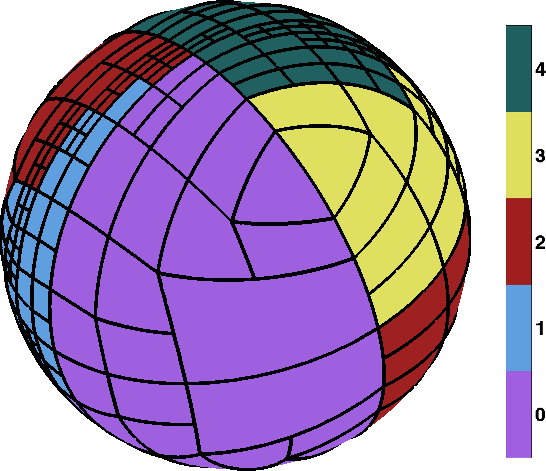}
    \end{minipage}
    \hspace{.05\columnwidth}
    \begin{minipage}[c]{.4\columnwidth}
    \begin{tabular}{r|r}
       $P$ & time int./step \\
      \hline
       16 & 1.115 \\
       32 & 0.557 \\
       64 & 0.297 \\
      256 & 0.092 \\
    \end{tabular}
  \end{minipage}
\end{center}
\caption{Left: Partition of the spherical Example 2 between five
  MPI processes, indicated by the color scale, at simulated time $T = 1$.
  The refinement is invariant under the parallel partition
  and follows the front, which is not aligned with the equator.  Right: Strong scaling
  for a uniform run at level 8.}
\label{fig:spherical}
\end{figure}
To showcase a multiblock connectivity, we include results for the spherical Example 2 in
\Fig{spherical}, together with a strong scaling table from 16 to 256 MPI ranks.

\section{Conclusion}

We have presented the integration of an MPI-based forest-of-octrees adaptive
meshing code, \pforest, with a numerical solver for hyperbolic conservation
laws, \clawpack/\manyclaw, that implements threaded parallelism for a single
compute patch.
We abstract an interface to the parallel meshing code in order to derive the
schedule of neighbor exchanges, averaging/interpolation, and time integration
on the local patches.
This approach naturally lends itself to MPI/thread hybridization.
Apart from its parallel scalability, we favor the presented strategy
for its modularity, encapsulation, and versality.  Future work
will continue to optimize the parallel neighbor exchange patterns
and investigate a generic handling of arbitrary multi-block geometries.

\section*{Acknowledgements}

We would like to thank the Texas Advanced Computing Center (TACC) for access to
the Stampede supercomputer under allocations TG-DPP130002 and TG-ASC130001
granted by the NSF XSEDE program.
The authors acknowledge valuable discussion with Randy LeVeque, Marsha
Berger, and Hans-Petter Langtangen.  We also acknowledge David
Ketcheson and the KAUST sponsored HPC$^3$ numerics workshop at which
the initial phases of this project were first discussed.  The second
author would like to also acknowledge the Isaac Newton Institute
(Cambridge, UK), where much of the preliminary development work for
\forestclaw was done.
The fourth author recognizes Simula Research Lab, Norway, for funding.
The leaf/patch paradigm was independently
presented by B.\ as part of a talk at the SCI Institute, Salt Lake City, Utah,
USA in July 2011.

\bibliographystyle{unsrt}

\end{document}